\documentstyle[multicol,aps,psfig,amsfonts,amssymb,prb]{revtex} 

\begin{document}
\draft
\preprint{{\bf ETH-TH/99-??}}

\title{Engineering Superconducting Phase Qubits}

\author{$^{a\,}$Gianni Blatter, $^{a,c\,}$Vadim B.\ Geshkenbein, 
and $^{b,c\,}$Lev B.\ Ioffe}

\address{$^{a\,}$Theoretische Physik, ETH-H\"onggerberg, CH-8093
  Z\"urich, Switzerland}

\address{$^{b\,}$Department of Physics and Astronomy, Rutgers University,
Piscataway, NJ 08854, USA}

\address{$^{c\,}$L.D.\ Landau Institute for Theoretical Physics,
  117940 Moscow, Russia}

\date{\today}
\maketitle
\begin{abstract}
\parbox{14cm} {The superconducting phase qubit combines Josephson
junctions into superconducting loops and defines one of the promising
solid state device implementations for quantum computing. While
conventional designs are based on magnetically frustrated
superconducting loops, here we discuss the advantages offered by
$\pi$-junctions in obtaining naturally degenerate two-level
systems. Starting from a basic five-junction loop, we show how to
construct degenerate two-level junctions and superconducting phase
switches. These elements are then effectively engineered into a
superconducting phase qubit which operates exclusively with switches,
thus avoiding permanent contact with the environment through external
biasing. The resulting superconducting phase qubits can be understood
as the macroscopic analogue of the `quiet'
$s$-wave--$d$-wave--$s$-wave Josephson junction qubits introduced by
Ioffe {\it et al.} [Nature {\bf 398}, 679 (1999)].}

\end{abstract}

\vspace{0.5truecm}
\vspace{-0.4truecm}

\begin{multicols}{2}
\narrowtext

\noindent
In a quantum computer the information is stored in quantum two-level
systems --- these qubits replace the familiar bits of the classical
computer as the basic computational unit. Calculations are performed
through the usual quantum state evolution, modifying the superposition
of basis states on individual qubits and the entanglement of states
between different qubits \cite{Ekert}.  Maintaining the coherence of
the quantum device throughout the calculation is the prime challenge
in the quest for quantum computation \cite{Haroche}: While the quantum
states have to be manipulated from outside in order to carry out the
specific computational task, the device should be maximally decoupled
from the environment in order to avoid decoherence. Proposals for
qubits based on trapped atoms \cite{Cirac-Zoller,Monroe}, photons in
QED cavities \cite{Turchette}, or nuclear spins (NMR)
\cite{Gershenfeld} are prime candidates for meeting these
requirements, however, it seems that upscaling to a useful computer
with of order 10$^4$ qubits is difficult. Here is where solid state
implementations of qubits enter the field, offering high manufacturing
variability as well as scalability based on highly developed nanoscale
technology. The large number of degrees of freedom associated with
solid state devices challenges the maintainance of coherence.  As of
today, this problem has been met by either resorting to well isolated
spins (on quantum dots \cite{Loss} or through deliberate doping of
semiconductors \cite{Kane}) or making use of the gapped quasi-particle
spectrum in superconducting structures
\cite{Schoen,Averin,SchoenN,Bosco,Ioffe,Feigelman,Mooij}. The
charge--phase (cooper-pair--vortex) duality in Josephson devices then
admits two classes of qubit implementations: The ``charge'' devices
\cite{Schoen,Averin,SchoenN} operate in the regime $E_C \gg E_J$
(where $ E_C = e^2/2C$ is the charging- and $E_J = I_c \Phi_0 / 2 \pi
c$ is the Josephson coupling energy; $C$ = capacitance, $I_c$ =
critical current, $\Phi_0 = hc/2e$ is the flux quantum),
distinguishing states through their charge, while the superconducting
phase qubits (SPQB) \cite{Bosco,Ioffe,Feigelman,Mooij} are based on
strongly coupled junctions with $E_J \gg E_C$. In ``phase'' qubits the
states are usually distinguished by the direction of a circulating
current in the Josephson loop, but recently a proposal for a `quiet'
qubit has been made \cite{Ioffe} which is effectively decoupled from
the environment. Here, we introduce the $\pi$-junction \cite{BKS}, a
Josephson junction with a ground-state characterized through a
$\pi$-phase shift across the contact, as a new building-block of
superconducting phase qubits and discuss how it can be used in the
design of stable and switchable phase qubits.

In the following, we first develop the main ideas leading to the qubit
design based on small-inductance multi-junction SQUID loops.  The
analysis of the robustness of these loops against static and dynamic
fluctuations in electric- and magnetic fields will then guide us in
the specific engineering of our five-junction loop, frustrated by a
$\pi$-junction and defining a perfectly degenerate two-level
system. We show how to construct a double-periodic $\pm\pi/2$-junction
from this five-junction loop and discuss its similarity with the
SD-Josephson junction introduced by Ioffe {\it et al.}
\cite{Ioffe}. We then concentrate on operational aspects of the
superconducting phase qubit: its efficient isolation from the
environment motivates the design and use of superconducting phase
switches and its trivial idle state is realized through a dynamically
decoupled degenerate two-level system. We end with a discussion of
possible implementations of $\pi$-junctions.

The classic phase qubit has been introduced in an early paper by Bocko
{\it et al.} \cite{Bosco} and involves a conventional two-junction
SQUID loop frustrated by an external magnetic field. The two states
used in the construction of the qubit are characterized by clock- and
counterclockwise circulating currents or by the corresponding phase
drops across the junctions. For quantum computing we are interested in
qubits with a minimal coupling to the environment, hence we wish to
have a situation where the phase is maximally decoupled from the flux,
with the two quantum states mainly distinguished by the phase
variable. Such a `quiet' phase qubit producing no currents in the
SQUID loop has been suggested recently \cite{Ioffe} and involves
doubly periodic Josephson junctions with minima characterized by the
phase variable $\phi = 0$ and $\phi = \pi$ (some unavoidable current
flow remaining in the junction area itself is discussed
below). However, these junctions are made from sandwiches combining
$s$- and unconventional $d$-wave superconductors and are difficult to
fabricate, hence alternative designs compromising with some residual
coupling between the phase and the flux are still highly welcome.
\begin{figure} [bt]
\centerline{\psfig{file=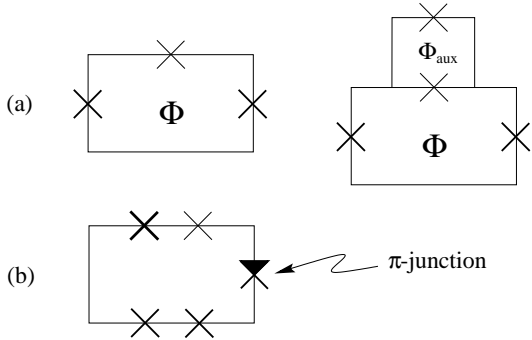,width=7cm,height=4.5cm}}
\narrowtext\vspace{4mm}
\caption{(a) The three junction SQUID studied by Mooij {\it et al.}
[15] where the two degenerate ground states with clock- and
counterclockwise circulating currents are realized at maximal
frustration with the penetrating flux $\Phi = \Phi_0/2$; changing the
flux $\Phi_{\rm aux}$ in the auxiliary SQUID loop produces an
effective junction with varying coupling, i.e., a tunable junction.
(b) Five-junction loop studied in the present paper. The frustration
is achieved through a $\pi$-junction.}
\end{figure}
Unfortunately, the conventional two-junction SQUID cannot offer a
solution: The constraint $2\pi\Phi/\Phi_0=\phi_1 + \phi_2$ relates the
total flux $\Phi$ through the loop with the gauge invariant phase
drops $\phi_1$ and $\phi_2$ across the junctions. For a large
inductance loop with $L I_c/c \gg \Phi_0$ the potential energy
$V(\phi_1,\phi_2;\Phi_{\rm ext} = \Phi_0/2) = E_J
[2-\cos\phi_1-\cos\phi_2] + (\Phi-\Phi_{\rm ext})^2/2L$ exhibits
pronounced minima of order $E_J$ but the states involve unfavorably
large fluxes of order $L I_c/c \sim \Phi_0$, easily leading to
magnetic crosstalk in an array of qubits. On the other hand, in a
small inductance loop, where $L I_c/c \ll \Phi_0$, the minima are
shallow, of order $E_J (L I_c/c\Phi_0)$. In order to effectively
decouple the phase variables from the flux we have to resort to small
inductance loops with three or more junctions. Here, both requirements
of well defined minima (of order $E_J$) and small fluxes ($L I_c/c \ll
\Phi_0$) can be satisfied simultaneously. A first proposal for a
superconducting phase qubit using a four-junction loop has been
discussed by Feigel'man {\it et al.} \cite{Feigelman} and a detailed
analysis of the three-junction loop involving a tunable third junction
has recently been given by Mooij {\it et al.} \cite{Mooij}, see Fig.\
1. In a small inductance three-junction loop (with $L I_c/c \sim
10^{-3} \Phi_0$) frustrated at $\Phi_{\rm ext} = \Phi = \Phi_0/2$, the
third junction is slaved to the other two, $\phi_3 = \pi - \phi_1
-\phi_2$, but the remaining two degrees of freedom are sufficient to
define pronounced minima ($\phi_1 = -\phi_2 \equiv \phi_{\rm min} =
\pm \pi/3$) in the potential energy (see Fig.\ 2)
\[
V(\phi_1, \phi_2) = E_J [3 - \cos\phi_1 - \cos\phi_2 -
\cos(\pi-\phi_1-\phi_2)],
\]
while at the same time the fields produced by these states are small
and hence do not perturb neighboring qubits. Quantum operation of the
loop is introduced through the finite junction capacitance $C$: with
$E_J/E_C \approx 25$ we have a plasma frequency $\omega_p \sim
\sqrt{E_J E_C} \sim E_J /5$ (placing the device into the semiclassical
regime with a well defined phase basis) and a tunneling gap $\Delta
\sim \hbar \omega_p \exp(-\sqrt{\alpha E_J/E_C})\sim 10^{-3} E_J$,
with $\alpha$ a numerical of order one ($\alpha = 8$ for a
conventional small capacitance Josephson junction
\cite{SchoenZaikin}). Single qubit operations are performed through
changes in the magnetic fluxes through the main (three-junction) loop,
$\Phi$, and the auxiliary (two junction) loop of the tunable third
junction, $\Phi_{\rm aux}$, see Fig.\ 1. Typical qubit operations are
carried out within the time scale $t_{\rm op}\sim \hbar/\Delta$ set by
the tunneling gap $\Delta$. The quantum entanglement between two
qubits can be modified through a tunable inductive coupling, where the
coupling loop is opened and closed by a tunable junction as realized
through a two-junction SQUID loop. First estimates of decoherence
rates for this type of device give favorable values \cite{Tian}.
\begin{figure} [bt]
\centerline{\psfig{file=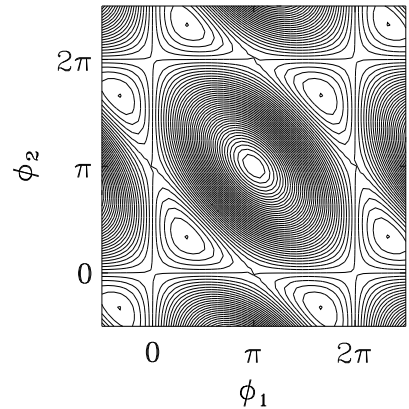,width=5.0cm,height=5.0cm}}
\narrowtext\vspace{8mm}
\caption{Contour plot of the potential energy
$V(\phi_1,\phi_2;\Phi_0/2)$ for the symmetric frustrated three
junction loop versus gauge invariant couplings $\phi_1$ and
$\phi_2$. The apparent difference in distance between the two
intracell- and intercell minima is due to the projection on the
$\phi_1$-$\phi_2$ plane.}
\end{figure}
{\it Charge stability:} While the decoupling between phase and flux in
a superconducting phase qubit can be achieved in small inductance
$n\geq 3$-junction loops, the decoupling between phase and random
offset charges in the environment can be realized by breaking the
symmetry of the loop and choosing one junction weaker than the
remaining ones \cite{Mooij}. E.g., in a three junction loop the state
$|+\rangle = |\phi_1=\phi_2=\phi_3= \pi/3\rangle$ can decay into the
state $|-\rangle$ through three different paths $\gamma_i$, $i=1,2,3$,
where a flux $2\Phi_0/3$ enters through junction $i$ and leaves
half-half through the other two, see Fig.\ 3 (due to the small
inductance $L$, $\Phi \approx 0$ on the scale of $\Phi_0$).  A
tunneling process from $|+\rangle$ to $|-\rangle$ then must be
calculated by adding the three amplitudes along $\gamma_i$ coherently,
$\Gamma_{+\to -} = |\sum_{i=1}^3\ _i\langle -|+\rangle|^2$.  The
presence of a charge $Q_i$ on the $i$-th island, e.g., due to random
offset charges, will then induce an additional Aharonov-Casher phase
$\exp(2\pi i Q_i/2e)$ between the two flux paths enclosing this
island, leading to unwanted changes in the tunneling amplitude
$\Gamma_{+\to -}(Q_i)$. Breaking the loop's symmetry with one weak
junction channels the large flux through this junction and spoils this
interference effect, rendering the qubit stable against random static
offset charges. Note that dynamically fluctuating charges still can
spoil the proper performance of the qubit through a time-like
Aharonov-Casher interference effect. In order to prevent such
dynamical modifications of the individual trajectory we have to
protect the qubit from fast changes in the offset charges due to
defects moving on a scale of the Rabi- or tunneling time
$\hbar/\Delta$.
\begin{figure} [bt]
\centerline{\psfig{file=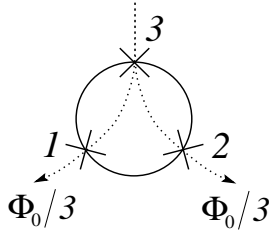,width=3.5cm,height=3.1cm}}
\narrowtext\vspace{4mm}
\caption{Reversing the current in the loop involves voltage pulses in
the junctions, which can be viewed as arising from a flux $2 \Phi_0/3$
entering through one junction and leaving half-half through the other
two junctions. The choice for the flux to enter through any one of the
three junctions leads to Aharanov-Casher type interference effects
which can sensitively influence the performance of the qubit.}
\end{figure}
{\it Flux stability:} While the phase qubit can be constructed to
resist charge fluctuations, it is very susceptible to any magnetic
field variation, as this is the prime external signal used in setting
up the two-level system and in manipulating the qubit. The requirement
on the field stability is quite nontrivial: Assume an imprecision
$\delta\Phi$ in the frustrating flux acting over $N_{\rm op}\approx
10^3-10^4$ operations (note that a device able to carry out $\sim
10^4$ operation can be run indefinitely using error correction
techniques \cite{Errorcorr}). In our current-phase decoupled loops
such a change $\delta\Phi$ in the flux will produce a shift of order
$(\delta\Phi/\Phi_0) E_J$ in the relative energy of the two states.
The total accumulated phase $\delta\phi\sim E_J(\delta\Phi/\Phi_0)t
/\hbar \sim (E_J/\Delta) (\delta\Phi/\Phi_0) N_{\rm op}$ should remain
small (on the scale of $2\pi$) and hence requires a precision
\[
\frac{\delta \Phi}{\Phi_0} \sim \frac{\Delta}{E_J N_{\rm op}} 
\sim 10^{-6} - 10^{-7}. 
\]
The equivalent noise analysis reads $\langle (\delta\phi)^2\rangle
\!\sim \!(E_J/\hbar \Phi_0)^2$ $\int dt_1\, dt_2 \, \langle
\delta\Phi(t_1)\, \delta\Phi(t_2)\rangle \sim (E_J t/\hbar)^2 \langle
\Phi_\omega^2\rangle / t \Phi_0^2 \sim 1$, resulting in a noise level
required to stay below
\[
\langle\Phi_\omega^2\rangle^{1/2} \sim \frac{\Phi_0}{E_J}
\sqrt{\frac{\hbar \Delta}{N_{\rm op}}} \sim \Phi_0 \times
[10^{-8}-10^{-9}]/\sqrt{{\rm Hz}},
\]
where we have assumed a typical coupling energy in the range
$E_J/\hbar \sim 10^2$ GHz and $t \sim N_{\rm op}\hbar/\Delta$ in the
MHz range (using $\Delta/E_J \sim 10^{-3}$ and $N_{\rm op} \sim
10^3$).

{\it $\pi$-junction:} A solution to this problem is offered by
replacing the external bias field through a $\pi$-junction in the
loop. This modification not only removes the need for a permanent and
quiet bias signal, it also facilitates manufacturing as requirements
on the uniformity of the loop size can be relaxed. $\pi$-junctions can
be constructed using crystals exhibiting unconventional
superconductivity, e.g., via placing the loop contacts normal to the
$\pm$ lobes of a $d$-wave superconductor as in the famous experiment
by Wollmann {\it et al.}  \cite{Wollman} demonstrating the $d$-wave
symmetry of the superconducting state in cuprate superconductors.  An
all $d$-wave small-inductance $\pi$-SQUID loop containing one $\pi$-
and one $0$-junction has recently been built and successfully tested
by Schulz {\it et al.} \cite{Schulz}.  Alternatives, such as
sandwiches of $s$-wave--ferromagnetic-metal--$s$-wave (SFS-junctions
\cite{BBP}) will be discussed below. Note that the $\pi$-junction acts
as a trivial static phase shifter in the loop and does not perform any
quantum action, hence requirements on its fabrication are not
stringent.
\begin{figure} [bt]
\centerline{\psfig{file=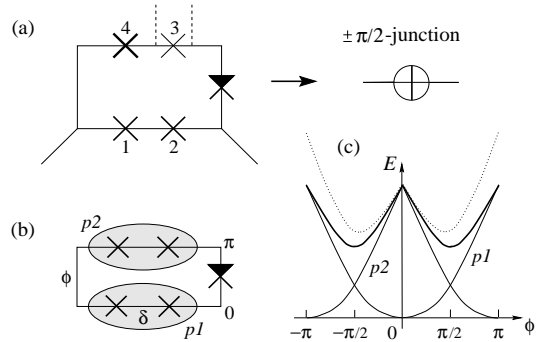,width=7cm,height=4.5cm}}
\narrowtext\vspace{4mm}
\caption{(a) Construction of an effective $\pm\pi/2$-junction from the
five-junction loop. The $\pm\pi/2$-junction has degenerate minima with
phase drops $\pm \pi/2$ across the external legs. In order to switch
the loop between its two states, a voltage pulse is applied to the
weakest junction (dashed lines). (b) Grouping the four junctions into
two pairs and minimizing the energy of each pair $p1$ and $p2$ allows
for an easy calculation of the energy-phase relation of the
$\pm\pi/2$-junction (c): the thin lines are the individual energies of
the two pairs with a relative shift by $\pi$, the thick line is their
sum for a symmetric situation producing minima at $\pm\pi/2$. Breaking
the symmetry (dotted line) produces degenerate minima away from
$\pm\pi/2$.}
\end{figure}
{\it Five-junction loop:} The above arguments then motivate the design
of the five-junction small-inductance SQUID-loop with one strong
$\pi$-junction as our basic device, see Fig.\ 1(b). Since the loop
cannot trap flux, the remaining four junctions are related via $\pi =
\phi_1 + \phi_2 + \phi_3 + \phi_4$. The choice with four junctions in
the loop is convenient (but not unique) as it allows for an easy
construction of sub-units performing specific tasks. In particular,
selecting the phase difference $\phi_1+\phi_2$ through appropriate
contacts we construct a $\pm\pi/2$-junction with minima at $\pm
\pi/2$, see Fig.\ 4(a). Choosing all junction couplings $E_i, i = 1,
\cdots, 4$ equal naturally defines a $\pm\pi/2$-junction (due to the
required charge stability we have to break the symmetry and thus have
to use the somewhat more complicated couplings $E_1=E_2=E_J$,
$E_3=(1-\gamma)E_J$, and $E_4=(1+\gamma')E_J$, $\gamma,\gamma' >
0$). It turns out, that the physics underlying the functionality of
this $\pm\pi/2$-junction is identical to that of the SD-Josephson
junction proposed earlier \cite{Ioffe}: Consider the idealized
five-junction loop shown in Fig.\ 4(b) where the four equal $s$-wave
junctions ($E_i = E_J$) have been arranged in pairs. Let us fix the
phase drop $\phi$ over the two first junctions and minimize the energy
of this pair, $E_{p1}(\phi) = {\rm min}_\delta E_J[2-\cos\delta-
\cos(\phi-\delta)]\rightarrow\delta=\phi/2$. The resulting
energy-phase relation $E_{p1}(\phi) = 2 E_J [1-|\cos(\phi/2)|]$ is
similar to that of a clean SNS-junction \cite{Ishii} with sharp cusps
at $\phi = \pm\pi$, see Fig.\ 4(c). Treating the second pair of
junctions in a similar way, a simple addition provides us with the
optimized energy-phase relation of the $\pm\pi/2$-junction,
$E_{\pm\pi/2}(\phi) = E_{p1}(\phi) + E_{p2}(\phi+\pi)$, where the
phase of the second pair is shifted by $\pi$ due to the presence of
the $\pi$-junction in the loop. For the symmetric setup with $E_{p1} =
E_{p2}$ we obtain minima at $\pm\pi/2$, while breaking the symmetry
will shift the minima away from $\pm\pi/2$ to values
$\pm(\pi/2-\alpha)$, see Fig.\ 4(c). It is then important to note that
while breaking the symmetry does shift the minima it {\it does not}
lift the degeneracy between the two minimal states, a crucial point in
the design of a qubit with a trivial idle state (in fact, this
degeneracy is a consequence of the time reversal symmetry in the
hamiltonian).

{\it {\rm SD}-Josephson junction versus five-junction loop:} Consider
then the SD-junction sketched in Fig.\ 5(a). We choose a contact with
a clean normal metal layer (N) providing the coupling between the
$s$-wave (S) and the $d$-wave (D) superconductor via the proximity
effect: electrons incident on the $s$-wave superconductor from the
left are reflected back as holes, a process known as Andreev
reflection \cite{Andreev}. In turn, the hole is reflected from the
$d$-wave superconductor as an electron and a phase-sensitive Andreev
state is formed carrying supercurrent across the normal metal
layer. Due to the $d$-wave symmetry, the Andreev levels naturally
split into two families contributing the coupling energies
$E_{f1}(\phi)$ and $E_{f2}(\phi+\pi)$, where the phase shift $\pi$ is
due to the sign change in the $d$-wave condensate under a rotation by
$\pi/2$, see Fig.\ 5(a). For a clean metal layer the couplings are
given by a sequence of parabolas $\propto E_J \phi^2/2$ with sharp
cusps at $\pm\pi$, see Ref.\ \onlinecite{Ishii}, and we obtain a
similar energy-phase relation as for the five-junction loop.  A
symmetry breaking with $E_{f1} \neq E_{f2}$ corresponds to a
misalignment of the SD boundary, producing a stronger coupling via one
of the families --- a misalignment in the $d$-wave crystal then leads
to a shift of the minima away from $\pm\pi/2$ but {\it does not}
destroy the degeneracy of the two-level system.
\begin{figure} [bt]
\centerline{\psfig{file=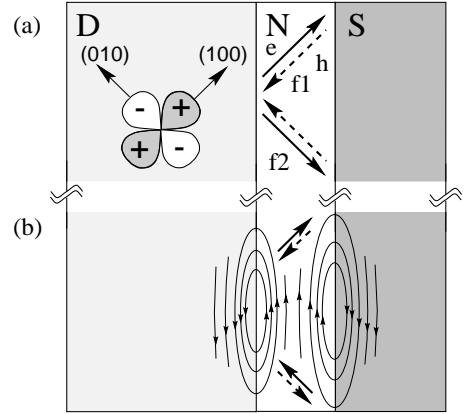,width=6cm,height=5.5cm}}
\narrowtext\vspace{4mm}
\caption{(a) Clean SND Josephson junction: the superconducting
coupling between the $s$- and $d$-wave superconductors is carried by
phase sensitive Andreev states. These coherent electron-hole states
form two families $f1$ and $f2$ with coupling energies shifted by
$\pi$ --- they correspond to the junction pairs $p1$ and $p2$ in Fig.\
4 (b) and produce an analogue energy-phase relation as for the
$\pm\pi/2$-junction, see Fig.\ 4(c). (b) The energy minima at $\phi =
\pm \pi/2$ carry a junction current flowing {\it along} the normal
metal layer and returning through the adjacent superconductors.}
\end{figure}
The obvious question to ask then is about the analogue of the
circulating currents in the five-junction loop for the SND-junction:
Indeed, as discussed by Huck {\it et al.} \cite{Huck}, the two
degenerate ground states at $\pm\pi/2$ are linked to a current flowing
up or down the intermediate metal layer and returning through the
superconductors on the side, thereby producing local magnetic fields
of size of the lower critical field $H_{c1} \approx
\Phi_0/4\pi\lambda^2$ decorating the metal layer on either side, see
Fig.\ 5(b) (here, $\lambda$ denotes the penetration depth; for a dirty
metal of thickness $d$ and with a mean free path $l$ the current flow
is suppressed by a factor $(l/d)^4$); hence the `quiet' qubit
advertised in Ref.\ \onlinecite{Ioffe} turns out to be less ideal then
originally expected, with some residual magnetic coupling surviving on
the level of the junction itself, a point which has been missed in
Ref.\ \onlinecite{Ioffe}. In an idealized situation with a symmetric
(back-) flow through the superconductors the fluxes trapped on the two
sides of the normal-metal layer would be of opposite sign and thus
would compensate one another, producing a magnetic far-field of
quadrupolar nature. However, in a SND junction we cannot expect the
backflow through the superconductors to be symmetric, hence the
dipolar component of the magnetic field is only partly
compensated. Still, the SND-junction teaches us a further trick in our
struggle to minimize the coupling of the qubit to the environment:
shaping the five-junction loop into a crossing double loop in the form
of an `8' produces a mutual compensation of the flux produced by the
circulating current. Note that this idea can only be realized when the
loop frustration is produced by a $\pi$-junction rather than an
external magnetic field.

{\it Qubits from five-junction loops and {\rm SD}-Josephson
junctions:} Given the five-junction loops and SD-Josephson junctions
above the qubit follows immediately from a simple
shift-operation. Rather than having the minima at $\pm\pi/2$, as is
the case for the five-junction loop and for the SD-Josephson junction,
we wish them placed at $0$ and $\pi$. Such a shift in the minima is
achieved by adding a second strong $\pi/2$-junction in series. This
strong $\pi/2$-junction with large coupling energy and large
capacitance acts as a classical shift-device and does not contribute
to the quantum evolution of the qubit. For the five-junction loop we
simple add a second, strongly coupled $\pi/2$-junction in series, see
Fig. 6(a), while in the case of the SD-Josephson junction a simple
termination with another DS'-junction defining a SDS'-Josephson
junction will do. In order to distinguish these double periodic
junctions with minima at $0$ and $\pi$ from the $\pm\pi/2$-junctions
we call them $2\phi$-junctions (alluding to the double periodicity)
and use the special symbol shown in Fig.\ 6(a).  The reason for adding
this $\pi/2$ phase-shift is an operational one: combining the
$2\phi$-junction together with a conventional $s$-wave junction into a
SQUID loop we can lift the degeneracy between the two ground states as
the $s$-wave junction is frustrated in the $\pi$-state --- this will
then allow us to define the phase-shift operation of the individual
qubits.
\begin{figure} [bt]
\centerline{\psfig{file=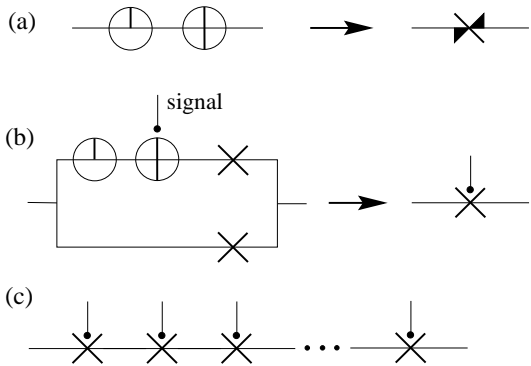,width=7cm,height=4.8cm}}
\narrowtext\vspace{4mm}
\caption{(a) Combining a strong (classical; left circle with upward
bar) $\pi/2$-junction and a (quantum) $\pm\pi/2$-junction produces the
$2\phi$-junction with minima shifted to $0$ and $\pi$.  (b) Switchable
$s$-wave junction (phase switch): switching the $\pm\pi/2$-junction
between $\pi/2$ and $-\pi/2$ the phase drops across the two $s$-wave
junctions either differ by $\pi$ or $0$. As a consequence, the total
coupling across the loop changes from $E_{sJ}^{\uparrow\downarrow} =
[-E_1+E_2]$ to $E_{sJ}^{\uparrow\uparrow} = [E_1+E_2]$.  (c) A
sequential array of $s$-wave switches allows to enhance the dynamical
range $E_{sJ}^{\uparrow\downarrow}/ E_{sJ}^{\uparrow\uparrow}$ of the
switch.}
\end{figure}
{\it Qubit manipulation:} Given the quantum state $|\Psi\rangle =
[|0\rangle + a \exp(i\chi) |1\rangle]/\sqrt{1+a^2}$ of the qubit,
single qubit operations serve to manipulate the phase $\chi$ and the
amplitude $a$ in the superposition (we call the corresponding
operations a `phase-shifter' and an `amplitude-shifter',
respectively).  A further point of criticism and potential improvement
then is in the manipulation of the qubit's quantum state. Again,
permanent connections to the qubit used to modify their relative
position in energy or the coupling between the two states are
potential carriers of noise.  One might then come up with the
following wireless scheme for qubit operation inspired by the usual
NMR technique: A qubit with an energy separation $\Delta$ of its
levels can be manipulated by means of resonant microwave pulses. A
deliberate variation of the coupling energies $\Delta_k$, $k =
1,\cdots N_{\rm qu}$, during the construction phase then permits to
address each qubit in an array individually by selecting the proper
eigenfrequency $\Delta_k/\hbar$ for the microwave signal. With $N_{\rm
qu}$ qubits in the array, the typical distance between resonances is
of order $\delta \Delta = \Delta /N_{\rm qu}$.  Assume we are
manipulating the $k$-th qubit by tuning the microwave signal to its
resonance frequency $\Delta_k/\hbar$. The transition time is related
to the $ac$-signal via $t_{\rm op}\sim h
/\sqrt{(\delta\varepsilon/2)^2 + |V_k|^2}$, where $V_k = \langle 0
|V|1\rangle_k$ is the matrix element of the perturbing $ac$-field and
$\delta\varepsilon$ denotes a possible deviation from resonance
\cite{LL}.  What is the probability to excite other qubits closeby in
energy? Such erroneous transitions appear with a probability
$|V_k|^2/2[(\delta\Delta/2)^2+|V_k|^2]$ and requiring them to be small
implies the condition $|V_k|^2 \ll (\delta \Delta /2)^2$.  This
translates into an operating time
\[
t_{\rm op} \gg N_{\rm qu} h /\Delta
\]
scaling linearly with the number $N_{\rm qu}$ of qubits in the system.
Hence, while upscaling the number of operations $N_{\rm op}$ puts
restrictions on the variation of the applied field, upscaling of the
number $N_{\rm qu}$ of qubits puts unfavorable limits on the operation
time $t_{\rm op}$ and renders the problem with keeping the flux
constant even more difficult.

{\it Switches:} An alternative scheme for a qubit manipulation with
minimal coupling to the environment is provided through
superconducting phase switches.  The basic idea derives from the
variable Josephson junction as realized through a small-inductance
two-junction SQUID loop: Frustrating the loop with a flux $\Phi_0/2$
produces a flat potential and hence a small effective coupling across
the loop. The problem with such a trivial implementation of the switch
is again with the use of an external bias current producing the
frustration in the {\it `off'} state.  Using our $\pm \pi/2$-junction
we can easily construct a `phase switch' which operates without
external bias field, see Fig.\ 6(b) and Ref.\ \onlinecite{Ioffe}. We
combine two $(\pm)\pi/2$-junctions and two conventional $s$-wave
junctions into a small inductance SQUID loop.  While the first
$\pi/2$-junction is set at $\pi/2$, leads attached to the second
$\pm\pi/2$-junction allow to switch between the states $\pm \pi/2$ via
the application of voltage pulses. If the second junction is in the
state $-\pi/2$, the two $\pi/2$-junctions compensate one another, the
couplings of the two $s$-wave junctions add, and the switch is closed.
On the other hand, if the second junction is set to $\pi/2$, the two
$s$-wave junctions compensate one another and the phase switch is
open. In formulae, consider the coupling $E_J(1-\cos \phi)$ of the
$s$-wave junction. With the two $\pi/2$-junctions compensating each
other one has an additive coupling $[E_{sJ,1} + E_{sJ,2}](1-\cos
\phi)$. On the other hand, with the $\pi/2$-junctions in parallel, the
associated $s$-wave junction is shifted to
$E_{sJ,1}[1-\cos(\phi-\pi)]$ and the total coupling is given (up to a
trivial constant) by the difference $[-E_{sJ,1} + E_{sJ,2}]
\cos\phi$. With this construction we have arrived at a switchable
$s$-wave junction or simply a superconducting phase switch.  Note that
the leads connecting the switch to the outside world only have to
bring a voltage-pulse to the switchable $\pm\pi/2$-junction and
therefore can be capacitivly decoupled from the switch loop, implying
an efficient cutoff of the low-frequency noise.

Unfortunately, we cannot hope to remove the coupling alltogether in
the open state of the switch, as this would require very precise
fabrication of identical junctions such that $E_{sJ}^{\uparrow
\downarrow} = -E_{sJ,1} + E_{sJ,2} = 0$. However, it is sufficient to
achieve a cancellation $E_{sJ}^{\uparrow \downarrow} \ll E_{sC}$, with
$E_{sC}$ the capacitive energy of the switch loop. Combining $n$
switch-loops into an array (see Fig.\ 6(c)) will then further reduce
the coupling by a factor $(E_{sJ}^{\uparrow \downarrow}/E_{sC})^{n-1}$
(this follows trivially from the analogy to a 1D-Hubbard chain with
hopping matrix element $t = E_{sJ}^{\uparrow \downarrow}$ and
interaction energy $U = E_{sC}$).

An attractive idea is to extend the above scheme to a switchable
$2\phi$-junction, combining two $(\pm)\pi/4$-junctions and two
$\pm\pi/2$-junctions into a SQUID loop (the operation follows the same
scheme as that of the superconducting switch with the
$(\pm)\pi/4$-junctions replacing the $(\pm)\pi/2$-junctions).  Such a
switchable $2\phi$-junction then would allow for a change in the
tunnelling gap through switching the barrier height separating the two
ground states, producing an efficient amplitude-shifter.
Unfortunately, small deviations in the precision of the junctions will
shift the minima of each $\pm\pi/2$-junction away from $\pm\pi/2$ (see
Fig.\ 4(c)), thereby removing the natural degeneracy of the two
ground-states in the combined $2\phi$-junction.  Though this still
allows to construct a working device, the trivial idle state will be
spoiled.

{\it Qubit functionality:} Let us finally combine the above elements
into a qubit and discuss its functionality in terms of single- and
two-qubit operations. We follow the scheme described previously in
Ref.\ \onlinecite{Ioffe}.  The qubit combines a $2\phi$-junction, a
switchable $s$-wave junction and a switchable large capacitance
$C_{\rm ext}$ into a SQUID-loop, see Fig.\ 7(a). A switchable
capacitance is easily realized by connecting a capacitance ($C_{\rm
ext}$) and a switchable $s$-wave junction (denoted by $s'$ in Fig.\
7(a) and in the following) in series: With a weak coupling
$E_{s'J}^{\uparrow \downarrow} < E_{C_{\rm ext}}$ the phase $\chi$ of
the island fluctuates freely and the capacitor is effectively switched
{\it off}. In order to switch {\it on} the capacitor $C_{\rm ext}$ we
have to establish a sufficiently strong coupling across the $s'$
junction, $E_{s'J}^{\uparrow \uparrow} > E_{C_{s'}}$, with
$E_{C_{s'}}$ the charging energy of the $s'$ junction, slaving the
phase $\chi$ of the island to the phase $\phi$ of the
$2\phi$-junction. In the idle state the $s$-wave junction is in the
{\it off} state (no coupling) and the capacitance is operating in
parallel with the $2\phi$-junction producing a large total capacitance
in the loop.  The resulting tunneling gap $\Delta_{\rm idle} \approx
\sqrt{E_J E_{C_{\rm ext}}} \exp[-(\alpha E_J/E_{C_{\rm ext}})^{1/2}]$
is small and the two degenerate ground states are dynamically
decoupled (the numerical $\alpha$ is of order 1 to 10 and depends on
the details of the potential barrier, $\alpha \approx 2.6$ for the
four junction loop; a switchable $2\phi$-junction would allow for a
potential decoupling of the two ground states). Note that this idle
state has a trivial state evolution with the relative amplitude and
phase of the quantum superposition remaining unchanged.  In order to
shift the relative phase $\chi$ in the superposition $|\psi\rangle =
[|0\rangle + a \exp(i\chi)|\pi\rangle]/\sqrt{1+a^2}$ the $s$-wave
junction is switched into the strong coupling state ({\it on}).  Its
frustration in the $|\pi\rangle$ state of the qubit removes the
degeneracy between the states $|0\rangle$ and $|\pi\rangle$. The time
evolution of the two-level system is given by the unitary matrix
$u_z(\varphi) = \exp(-i\sigma_z \varphi/2)$ with $\varphi = -2
E_{sJ}^{\uparrow \uparrow} t/\hbar$ and $\sigma_z$ the usual Pauli
matrix. Keeping the $s$-wave junction {\it on} during the time $t$
will shift the phase by $\chi = -2E_{sJ}^{\uparrow \uparrow}t/\hbar$.
Finally, the amplitude $a$ is modified by switching {\it off} the
large capacitance $C_{\rm ext}$ in the loop. With the capacitance now
determined by the small capacitance $C_q$ of the qubit (plus a
residual small capacitance from the switch) the tunneling gap $\Delta
\approx \sqrt{E_J E_{C_q}} \exp[-(\alpha E_J/E_{C_q})^{1/2}]$ is large
and the qubit exhibits Rabi oscillations with a frequency $\omega$ =
$2\Delta/\hbar$.  The corresponding time evolution is given by the
unitary matrix $u_x(\vartheta) = \exp (-i\sigma_x \vartheta/2)$ with
$\vartheta = - 2 \Delta t /\hbar$. Starting from $|\psi\rangle =
|0\rangle$ and keeping the external capacitance {\it off} over a time
$t$ will shift the amplitude $a=0$ to $a = \tan^2 \vartheta$.

Two-qubit operations are carried out using a similar scheme
\cite{Ioffe}: As two qubits are coupled into a SQUID loop by closing
an $s$-wave switch, the two states $|0, \pi \rangle$ and $|\pi,
0\rangle$ are frustrated and are separated from the states $|0,
0\rangle$ and $|\pi, \pi \rangle$ by the energy $2E_{sJ}^{\uparrow
\uparrow}$. This allows us to shift the phases of the states $|0, \pi
\rangle$ and $|\pi, 0\rangle$ relative to $|0, 0\rangle$ and $|\pi,
\pi \rangle$ and the two-qubit state evolves in time following the
evolution
\begin{eqnarray}
{\cal U}_{\rm ps} (\chi) =
\left(\begin{array}{cc}\!
u_z(\chi) & 0
\! \\ \!
0 & u_z(-\chi)
\! \end{array}\right)
\label{phase_shifter}
\end{eqnarray}
with $\chi = -2E_{sJ}^{\uparrow \uparrow}t/\hbar$. Combining this
`phase shifter' with several single-qubit operations $U_{i\mu}
(\theta)$ rotating the qubit $i$ by an angle $\theta$ around the axis
$\mu$ [$u_\mu(\theta) = \exp(-i\sigma_\mu\theta/2)]$ then allows us to
construct the nontrivial controlled NOT gate
\begin{eqnarray}
U_{\rm\scriptscriptstyle CNOT} &=& \exp(-i\pi/4) U_{2y}(\pi/2)
U_{1z}(-\pi/2) U_{2z}(-\pi/2) \nonumber\\ &\quad\cdot& U_{\rm
ps}(\pi/2) U_{2y}(-\pi/2).
\label{reconstruct_CNOT}
\end{eqnarray}

The proper operation of these qubits over $N_{\rm op}$ operations puts
some constraints on the dynamical range of the switch: We require that
the accumulated phase due to leaking in the open state is small,
\[
N_{\rm op} t_{\rm op} E_{sJ}^{\uparrow \downarrow} / \hbar < 1,
\]
hence $E_{sJ}^{\uparrow\downarrow}< \Delta / N_{\rm op}$. On the other
hand, we wish the phase shift operation to be equally fast as the
amplitude shift, hence $E_{sJ}^{\uparrow \uparrow}\gtrsim \Delta$. The
two conditions then result in the dynamical range $E_{sJ}^{\uparrow
\uparrow}/E_{sJ}^{\uparrow \downarrow} \gtrsim N_{\rm op}\sim 10^3$,
which can be achieved through the switch array described above, see
Fig.\ 6(c).
\begin{figure} [bt]
\centerline{\psfig{file=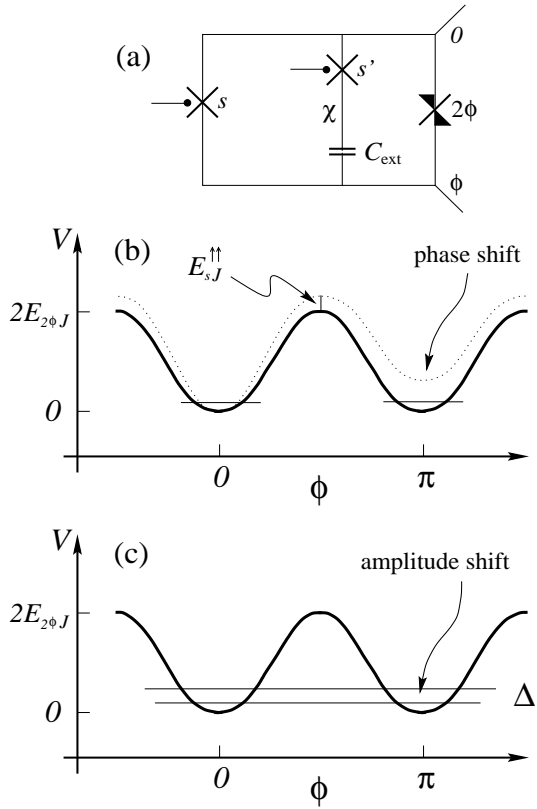,width=7cm,height=10.6cm}}
\narrowtext\vspace{4mm}
\caption{(a) Qubit made from a $2\phi$-junction, a switchable large
capacitance $C_{\rm ext}$, and a switchable $s$-wave junction.  (b)
Idle-state: the capacitance $C_{\rm ext}$ is switched in parallel to
the $2\phi$-junction blocking Rabi oscillations --- the
$s$-wave-junction is switched {\it off} to keep the states degenerate.
Phase-shift: The $s$-wave junction is {\it on} and its strong coupling
$2E_{sJ}^{\uparrow \uparrow}$ removes the degeneracy of the two
levels. (c) Amplitude-shift: The external capacitance is switched {\it
off} allowing the $2\phi$-junction to execute Rabi-oscillations.}
\end{figure}

{\it Realization of $\pi$-junctions:} Today, the most simple way to
realize a $\pi$-junction makes use of a copper-oxide $d$-wave
superconductor: connecting two $s$-wave superconductors (e.g., Pb) to
the $(1,0,0)$ and $(0,1,0)$ surfaces of a $d$-wave material (e.g.,
YBa$_2$Cu$_3$O$_7$), the sign change in the wave function of the
Cooper-pair produces the desired sign change or $\pi$-shift
\cite{Wollman}. However, long before the discovery of the high-$T_c$
copper-oxides, suggestions have been made for the construction of
$\pi$-junctions making use of magnetic impurities or bulk
ferromagnetic interlayers: Bulaevskii, Kuzii, and Sobyanin \cite{BKS}
have been the first to note that introducing magnetic impurities into
a tunneling junction leads to a sign change in the Josephson critical
current if the tunneling channel through the magnetic impurities is
strong enough to outplay the direct tunneling path. Later,
critical-current oscillations have been predicted to occur in clean
\cite{BBP} and dirty \cite{BBK}
superconductor--ferromagnet--superconductor (SFS) junctions
\cite{DAB}: Consider a Cooper-pair injected into a metallic layer
under an angle $\theta$ with respect to the interface normal ${\bf
n}$.  In a normal metal such a pair involves the states $|{\bf
p}\!\!\,\uparrow \rangle$, $|-{\bf p}\!\!\,\downarrow\rangle$. In a
clean ferromagnet, the energies of the states with equal momenta $p$
and opposite spins are split by the exchange field, $E_{p,\pm} =
\epsilon_p \pm E_{\rm xc}$.  Conservation of energy and momentum
parallel to the SF interface implies that the perpendicular momenta of
the two electrons are shifted in order to compensate for the different
exchange energy, $\delta {\bf p_n} = \pm E_{\rm xc} {\bf
n}/(v_{\rm\scriptscriptstyle F} \cos\theta)$, the sign of the shift
depending on the alignment of the spin with the exchange field. Thus
in the ferromagnet the pair involves the states $|[{\bf p}+\delta{\bf
p_n}] \!\!\,\uparrow\rangle$, $|-[{\bf p}-\delta{\bf p_n}]
\!\!\,\downarrow \rangle$ and carries a center of mass momentum ${\bf
P}_{\rm xc}=2 E_{\rm xc}{\bf n}/ (v_{\rm\scriptscriptstyle
F}\cos\theta)$, resulting in a spacially oscillating pair wave
function $\propto \cos ({\bf P}_{\rm xc} \cdot {\bf R}/\hbar)$ (here,
${\bf R}$ denotes the center of mass coordinate of the pair). The
proper integration over angles finally produces an oscillating
dependence of the critical current in the parameter $d E_{\rm
xc}/\hbar v_{\rm\scriptscriptstyle F}$, where $d$ is the thickness of
the ferromagnetic layer. For a dirty ferromagnet the relevant
parameter is $\sqrt{d^2 E_{\rm xc}/\hbar v_{\rm\scriptscriptstyle
F}l}$, with $l$ the mean free path of the dirty magnetic metal
\cite{BBK}. Experimentally, $T_c$ variations have been observed in
Nb/Gd multilayers \cite{Jiang} with varying thickness of the
ferromagnetic Gd layers, providing evidence for a $\pi$-coupling
realized at a thickness $d_{\rm Gd}\approx 20$~\AA.  Direct
observation of a sign change in the critical current at a definite
temperature and ferromagnetic-layer thickness has been recently
reported by Veretennikov {\it et al.} \cite{Veretennikov} in their
study of Nb-Cu/Ni-Nb junctions.

{\it $2\phi$-junctions:} Another basic idea to be mentioned in the
present context is the possibility to construct a `microscopic'
$2\phi$-junction (in contrast to the `macroscopic' $2\phi$-junctions
made from five-junction loops discussed above). Such junctions can be
obtained by an accurate quenching of the first harmonic coupling,
leaving the second harmonic with an energy $E_{\rm\scriptscriptstyle
J}(\phi) \propto \cos(2\phi)$ as the leading term.  A particular
scheme for the implementation of such a junction has already been
mentioned above, the SDS'-junction introduced in Ref.\
\onlinecite{Ioffe}; exploiting the node in the $d$-wave state, the
elimination of the first harmonic is achieved by balancing the two
lobes of opposite sign in wave function of the Cooper pair.

Another idea is to quench the lowest harmonic via disorder: Using a
dirty SF$_{\rm\scriptscriptstyle D}$S junction, the averaging over the
phase factors $\exp(i{\bf P}_{\rm xc} \cdot {\bf R})$ in the Cooper
pair wave function produce an efficient (i.e., exponential)
suppression in the Josephson coupling.  Within a non-interacting
electron description a similar suppression renders the second and
higher harmonics small, too. However, including effects of interaction
in the calculation of the second harmonic coupling we can trade these
exponential factors for an interaction vertex and arrive at a
significant second order coupling $E^{(2)}_{\rm\scriptscriptstyle J}
\sim E_0 (\lambda/ k_{\rm\scriptscriptstyle F}d)$, with $E_0 \equiv
k_{\rm\scriptscriptstyle F}^2 {\cal A} \hbar v_{\rm\scriptscriptstyle
F}/d$ the coupling strength of a usual metallic (SNS) junction and
$\lambda$ the dimensionless interaction parameter ($\lambda = VN_0$,
$V$ the interaction and $N_0$ the density of states; in obtaining
these estimates we have mimicked the disorder by a rough interface and
have applied estimates valid on short scales). In mesoscopic devices,
care has to be taken regarding fluctuation effects; finite mesoscopic
fluctuations in the first-order coupling can effectively compete with
the second harmonic, see Ref.\ \onlinecite{Spivak}.  A specific
discussion of a mesoscopic SF$_{\rm\scriptscriptstyle D}$S junction
exhibiting a double-periodic energy--phase relation (a mesoscopic
dirty SF$_{\rm\scriptscriptstyle D}$S $2\phi$-junction) has recently
been given by Zyuzin and Spivak \cite{ZyuzinSpivak}.

In conclusion, we have discussed the main basic principles involved in
the construction of superconducting phase qubits. We have shown that
this device family offers unique opportunities in the construction of
a trivial idle state, in removing dangerous couplings to the
environment, and the possibility to use phase switches rather than
permanently attached bias leads in the manipulation of the qubit. The
basic idea of the present work is that one can obtain a macroscopic
doubly-periodic or $2\phi$-junction device using five-junction loops
comprising a $\pi$-junction. Alternatively, a meso- or microscopic
implementation of such $2\phi$-junctions involving SDS' or dirty SFS
junctions opens up the possibility to design and develop a compact
superconducting phase qubit.

Financial support from the Swiss National Foundation via the `Zentrum
f\"ur theoretische Studien' is gratefully acknowledged.

\end{multicols}

\end{document}